\documentclass[pra,twocolumn,showpacs,superscriptaddress]{revtex4}
\usepackage{graphicx}

\def\nbZ{{\mathchoice {\hbox{$\sf\textstyle Z\kern-0.4em Z$}} 
{\hbox{$\sf\textstyle Z\kern-0.4em Z$}} {\hbox{$\sf\scriptstyle Z\kern-0.3em Z$}} 
{\hbox{$\sf\scriptscriptstyle Z\kern-0.2em Z$}}}}

\begin{document}

\title{Entanglement dynamics in the Lipkin-Meshkov-Glick model}
\author{Julien Vidal}
\author{Guillaume Palacios}
\author{Claude Aslangul}
\affiliation{Groupe de Physique des Solides, CNRS UMR 7588,Campus Boucicaut, 
140 rue de Lourmel, 75015  Paris, France}

\begin{abstract}
The dynamics of the one-tangle and the concurrence is analyzed in the Lipkin-Meshkov-Glick model which
describes many physical systems such as the two-mode Bose-Einstein condensates. We consider
two different initial states which are physically relevant and show that their entanglement dynamics are very different. A
semiclassical analysis is used to compute the one-tangle which measures the entanglement of one spin
with all the others, whereas the frozen-spin approximation allows us to compute the concurrence using
its mapping onto the spin squeezing parameter.

\end{abstract}

\maketitle

%
%
%
\section{Introduction}
%
%
%

In the recent years, the interplay between entanglement and quantum phase transitions  has been the subject of many
studies. Since the pioneering works on the Ising model in a transverse magnetic field \cite{Osborne02,Osterloh02} exhibiting
the key role played by entanglement in quantum critical phenomena, lots of effort have been devoted to
characterize the intricate structure of the ground state in spin systems
\cite{Latorre03,Latorre04_1,Gu03_1,Korepin03,Glaser03,Bose02,Syljuasen03_1,Syljuasen03_2,Verstraete04_1,Verstraete04_2,
Jin04,Roscilde04,Stauber04,Stelmachovic04,Alcaraz03,Osenda03,Fan04,Wellard04,Lambert04_1,Lambert04_2} as well as in electron
models
\cite{Gu03_2,Li04,Shi04}.  All these studies show that entanglement, a genuine quantum property, can be used to probe the
phase diagram of a  system and especially to detect quantum phase transitions.
Unfortunately, such analyses are often restricted to one-dimensional (1D) systems where exact solutions exist and allow
one to deal with the thermodynamical limit. Usually, higher-dimensional systems require a numerical treatment which is limited
by the Hilbert space dimension which grows exponentially with the number of degrees of freedom. 
However, when the Hamiltonian has a large symmetry group as is here the case, it is still possible to study the large-$N$ limit. 

In the present paper, we consider $N$ mutually interacting spins $1/2$ embedded in a magnetic field. The restriction of this
model to the fully symmetric subspace can be mapped, via the Schwinger representation, onto interacting bosons 
in a two-level system coupled by a tunneling term. Thus, although it has initially been introduced by Lipkin, Meshkov, and
Glick (LMG)
\cite{Lipkin65,Meshkov65,Glick65} in nuclear physics, this model is also relevant to describe the Josephson effect, or two-mode
Bose-Einstein condensate (BEC). This ubiquity in very different domains is certainly one of the reason why it has been
periodically rediscovered (see, for example, \cite{Botet82,Botet83,Cirac98}). Its integrability has even recently been proved in
a series of paper (see Ref.
\cite{Dukelsky04} for a review) and a complete solution has been derived using the algebraic Bethe ansatz \cite{Links03}. 
The ground state entanglement properties of this model have been
analyzed for both  ferromagnetic \cite{Vidal04_1,Dusuel04_3} and antiferromagnetic coupling \cite{Vidal04_2} using
the concurrence
\cite{Wootters98} (see below for details). Here, our goal is to analyze the entanglement dynamics which has, so far, been
mostly investigated in 1D spin systems \cite{Amico04,Montangero03}. 

The possibility of experiments in BEC's has led several groups to study the dynamics in the LMG model but entanglement has
been discussed only recently \cite{Micheli03,Hines03,Ng03,Ng04}. In this study, we investigate the time evolution of two
peculiar states which are relevant in the BEC context since they correspond to situations where all atoms are either in one of
the two modes or equally distributed between both modes. For these two states, we discuss the entanglement dynamics through two
observables: namely, the one-tangle which measures the entanglement of one spin with all others and the concurrence which
quantifies the two-spin entanglement. In the next section, we introduce the LMG model and discuss its phase
diagram. In Sec. \ref{Measures}, we define the entanglement measures used throughout this paper and give several canonical
examples.  The ground-state entanglement properties \cite{Vidal04_1,Dusuel04_3,Vidal04_2} are briefly recalled in
Sec. \ref{ground}. The quantum dynamics is presented in Sec. \ref{dynamics} for the two states
mentioned above. We show that the one-tangle and the concurrence have completely different behaviors depending
on the initial state. These characteristics can be understood in several limiting cases that we discuss in detail. 
%
%
%
\section{The Lipkin-MeshKov-Glick model}
\label{model}
%
%
%
We consider a system of mutually interacting spins $1/2$ embedded in a magnetic field described by the following 
Hamiltonian introduced by LMG \cite{Lipkin65,Meshkov65,Glick65} in a more general form:
%
%
\begin{eqnarray}
H&=&-\frac{\lambda}{N}\sum_{i<j} \sigma_{x}^{i}\sigma_{x}^{j}
 -h \sum_{i}\sigma_{z}^{i} \\
&=&-\frac{2 \lambda}{N} S_x^2 -2 h S_z + {\lambda \over 2},
\end{eqnarray}
%
%
where the $\sigma_{\alpha}$'s are the Pauli matrices and  $S_{\alpha}  =\sum_{i} \sigma_{\alpha}^{i}/2$. 
The prefactor $1/N$ is necessary to get a finite free energy per spin in the  thermodynamical limit.  
The Hamiltonian $H$ preserves the magnitude of the total spin and does not couple states having a different parity of
the number of spins pointing in the magnetic field direction (spin-flip symmetry): namely,
%
%
\begin{eqnarray}
\left[ H,{\bf S}^2 \right]&=&0,\\
\bigg[ H,\prod_i \sigma_z^i \bigg] &=&0.
\label{phaseflip}
\end{eqnarray}
%
%

The phase diagram of this model can be easily derived using a semiclassical description \cite{Botet83}. For
a ferromagnetic coupling ($\lambda>0$), the system undergoes a second-order quantum phase transition at the critical field 
$h_c= \lambda$, whereas for antiferromagnetic interactions, a first-order quantum phase transition occurs at zero field.

%
%
%
\section{Entanglement measures}
\label{Measures}
%
%
%

To analyze the entanglement dynamics, we focus here on two commonly used quantities: the one-tangle which measures the
entanglement of one spin with the others and the concurrence which measures the two-spin entanglement. 
For a given pure state $|\psi \rangle$, both quantities are computed from the density matrix 
$\rho=|\psi \rangle \langle \psi|$.

The one-tangle is defined as follows:
%
%
\begin{equation}
\tau_i=4 \det \rho_i^{(1)}=1-\sum_{\alpha} \langle \sigma_{\alpha}^{i}\rangle^2,
\label{taudef}
\end{equation}
%
%
where $\rho_i^{(1)}$ is the one-spin reduced density matrix obtained from $\rho$ by tracing out over all spins except
spin $i$ and where  $\langle A\rangle= {\rm Tr} \left[ \rho_i^{(1)} A \right]$. 
The one-tangle $\tau_i$ ranges between $0$ and $1$.
It vanishes  for a state such that $|\psi \rangle=|\phi \rangle_i \otimes |\varphi \rangle$ and reaches $1$ for
maximally entangled states such as the famous Einstein-Podolsky-Rosen (EPR) state \cite{Einstein35} 
%
%
\begin{equation}
\displaystyle{| {\rm EPR} \rangle={1
\over \sqrt {2}} 
\left(|\uparrow \downarrow\rangle- |\downarrow \uparrow \rangle \right)}.
\end{equation}
%
%

To measure the two-spin entanglement, we consider the concurrence $C$ introduced by Wootters \cite{Wootters98}. This quantity
is computed from the two-spin reduced density matrix $\rho_{i,j}$ obtained by tracing out $\rho$ over all spins except
spins $i$ and $j$. Next, we introduce the spin-flipped density matrix 
$\tilde \rho_{i,j}=\sigma_{y}\otimes\sigma_{y}\:\rho^{\ast}_{i,j}\:\sigma_{y}\otimes\sigma_{y}$ where 
$\rho^{\ast}_{i,j}$ is the complex conjugate of $\rho_{i,j}$. The concurrence $C$ is then defined by
%
%
\begin{equation}
C_{i,j}=\max\left\{ 0,\mu_{1}-\mu_{2}-\mu_{3}-\mu_{4}\right\} \equiv \max \{0,C^*_{i,j} \},
\label{concdef}
\end{equation}
%
%
where the $\mu_{k}$'s are the square roots of the four real eigenvalues of $\rho_{i,j} \: \tilde{\rho}_{i,j}$ and where
$\mu_{i} \geq \mu_{i+1}$. The
concurrence vanishes, for example, for any state $|\psi \rangle=|\chi \rangle_i \otimes |\phi \rangle_j
\otimes|\varphi \rangle$  and reaches its maximum value $C=1$ for states such as $| {\rm EPR} \rangle$.
 
Let us also mention that $\tau$ and $C$ are related through the Coffman-Kundu-Wootters (CKW) conjecture \cite{Coffman00}
stating that
%
%
\begin{equation}
\tau_i\geq \sum_{j\neq i}  C_{i,j}^2.
\label{CKWconj}
\end{equation}
%
%
Note that for pure two-spin states, we have an exact equality.
In the present study, we have systematically checked that this conjecture (\ref{CKWconj}) was always verified.

Finally, we would like to stress that tracing out the density matrix $\rho$ can induce or destroy some correlations 
between spins. For instance, a nonseparable state, such as the Greenberger-Horne-Zeilinger (GHZ) state \cite{Greenberger89} 
%
%
\begin{equation}
\displaystyle{| {\rm GHZ} \rangle=|\psi \rangle={1 \over \sqrt {2}} 
\left(|\uparrow \uparrow \uparrow\rangle+ |\downarrow \downarrow \downarrow\rangle \right)},
\end{equation}
%
%
has a maximum one-tangle $(\tau=1)$ and a vanishing concurrence $(C=0)$. It is thus important to keep in mind that these
measures only give partial information about the entanglement of the state.

%
%
%
\section{Ground-state entanglement}
\label{ground}
%
%
%
We recall in this section the main features of the ground state entanglement and we refer the reader to 
Refs. \cite{Vidal04_1,Dusuel04_3,Vidal04_2} for a detailed discussion. 

As explained in Sec. \ref{model}, the ground state of $H$ is well described by a mean-field approach showing that in the
thermodynamical limit the ground state is given by a product state \cite{Botet83} for which $\tau=C=0$.
However, in the symmetric phase ($\lambda>0$ and $h<\lambda$), we underline that the ground state is twofold degenerate  in
the thermodynamical limit  whereas at finite-$N$, it is unique (except for $h=0$). In the degenerate case, one must consider
the thermal density matrix at zero temperature \cite{Vidal04_1} to compute the entanglement properties. This implies
in particular that the one-tangle in this phase is given by
$\tau=1-(h/\lambda)^2$, and that $C=0$ in the large-$N$ limit.
However, the finite $N$ corrections to this separable asymptotic form allow one to capture a nontrivial behavior of the
rescaled concurrence $C_R=(N-1) C$. This rescaling is actually the coordination number of each site and takes into account the
fact that the two-site entanglement, as measured by the concurrence, is, in this model, equally shared between all sites. 

\begin{figure}[ht]
\includegraphics[width=100mm]{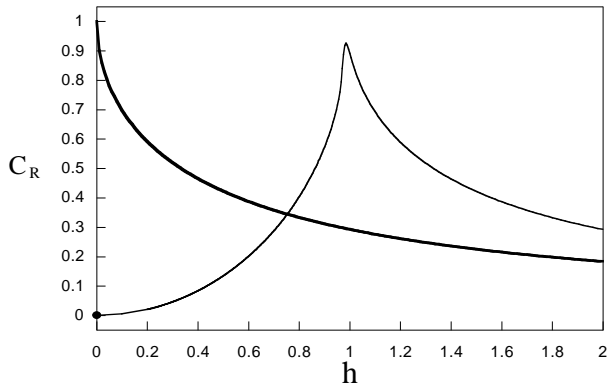}
\vspace{-15mm}
\caption{Rescaled concurrence of the ground state for $\lambda=+1$ (thin line) and
$\lambda=-1$ (thick line) ($N=10^3$ spins). In both cases, $C_R=0$ for $h=0$.}
\label{concGS}
\end{figure}

For a ferromagnetic coupling ($\lambda>0$), $C_R$ displays a cusplike behavior at the critical point $h_c=\lambda$ 
(see Fig. \ref{concGS}) where $C_R$ goes to $1$ in the thermodynamical limit. 

In the antiferromagnetic case, $C_R$ is a smoothly decreasing function of $h$ but is discontinuous at $h=0$ where a
first-order transition occurs (see Fig. \ref{concGS}). For $h=0^+$, $C_R$ reaches $1$ in the thermodynamical limit, but for
$h=0$, the ground state is given by zero total spin states for which $C_R=0$.

%
%
%
%

Except for this very special point, the ground state always lies in the maximum spin sector $S=N/2$ for any $\lambda$.
Moreover, at finite $N$ and in the large-field limit, the ground state is the fully polarized state in the $z$ direction,
%
%
\begin{equation}
|\psi_0 \rangle=|Z\rangle=\otimes_{i=1}^N |\uparrow \rangle,
\end{equation}
%
%
for which one also has $C_R=0$.

In the antiferromagnetic case, the entanglement properties of the ground state (restricted to the maximum spin sector
$S=N/2$) have also been analyzed by Hines {\it et~al.} \cite{Hines03} through the Von Neumann entropy $E(h)$ for bipartite
systems. In BEC language, the two subsystems correspond to the two components of the condensate, {\it i.~e.} to the states
polarized in the $\pm x$ direction. 
This quantity has thus nothing to do with the concurrence and its behavior is indeed very different: 
$E$ vanishes for $h=0$ (since $|\psi_0 \rangle=|X \rangle$) whereas $C_R$ is maximum (in this spin sector) and $E$ is maximum
in the large-field limit whereas $C_R$ vanishes. 

More recently, another entropy $S(h)$ has been analyzed in the LMG model \cite{Latorre04_3}. This entropy is based on a
bipartite decomposition in two subsytems of size $L$ and $N-L$. Here, $S(h)$ has been found to be maximum at the critical point
$h_c$ in agreement with a recent conjecture proposed by Hines {\it et al.} \cite{Hines04}. Moreover, a logarithmic scaling has
been observed and may reveal an underlying 1D conformal field theory \cite{Latorre04_3} for the LMG model. 

%
%
%
\section{Quantum dynamics}
\label{dynamics}
%
%
%

In this section, we analyze the quantum dynamics of two initial states belonging to the maximum spin subspace $S=N/2$: 

\begin{enumerate}

\item  the state $|X\rangle$ fully polarized in the $x$ direction which corresponds, in BEC language, to a state where
the  atoms are all localized in one of the two modes: 

\item the state $|Z\rangle$ fully polarized in the $z$ direction which corresponds, in the BEC language, to a state where
the atoms are equally distributed between both modes (catlike state).

\end{enumerate}

The quantum dynamics of these two states have already been widely studied for this model in the BEC context
\cite{Law01,Milburn97,Hines03,Micheli03,Gordon99,Zhang03,Kalosakas03_1,Kalosakas03_2}. Here, we focus on the entanglement
dynamics by studying the time evolution of $\tau(h,t)$ and $C_R(h,t)$, which do not depend on the spins kept in the trace
operation since states belonging to the subspace $S=N/2$ are invariant under the permutation group. This implies in particular
that
$\tau$ and $C_R$ only depend on the magnetization $\langle S_\alpha \rangle$ and on the correlations functions $\langle S_\alpha
S_\beta
\rangle$
\cite{Wang02}.

%
%
\subsection{Time scales}
%
%
As explained in Ref. \cite{Bernstein90}, the spectrum of $H$ remains discrete in the thermodynamical limit; {\it i.~e.} ,the
mean level spacing remains finite when $N$ goes to infinity. For equidistant levels, we would expect to have a Poincar\'e time
that does not depend on $N$. However, it turns out that the Poincar\'e time scales here linearly with $N$. As noticed by
Milburn {\it et al.} \cite{Milburn97}, the semiclassical dynamics (see below) mimicks very well the quantum dynamics up to a
given time scale $t_{sc}$ which is of the order of the Poincar\'e time. For times smaller than
$t_{sc}$, the system thus remains nonentangled and in the large-$N$ limit, it is described by the classical equations of
motion (see Appendix \ref{semiclassical}). For times larger than $t_{sc}$ quantum effects become important, some revivals of
the wavepacket are observed,  and the latter approach fails. It is thus in this regime that interesting and nontrivial
entanglement properties must be analyzed. Consequently, we always consider a dimensionless rescaled time $\mu=4 \lambda t/N$
which allows us to investigate the dynamics and to easily compare the results for various $N$. Of course, we have checked in
each situation that the time-averaged results were weakly sensitive to the the time range chosen provided this latter was
sufficiently large. 

We must also mention recent studies \cite{Kalosakas03_1,Kalosakas03_2} which reveal multiple time scales in the LMG
model (called the boson-Hubbard dimer model in these papers). However, the range of validity of these results is
confined to $h \ll \lambda/N$ (small-tunnelling-amplitude regime in boson language) which is not the regime
considered here. Yet we cannot exclude larger characteristic time scales but such a study is beyond the scope of the present
paper and we shall leave this question apart in the following. 

%
%
\subsection{The $|X \rangle$ state}
\label{xstate}
%
%

It is clear that the entanglement properties of the $|X \rangle$ state are insensitive to the transformation 
$h \leftrightarrow -h$. Thus, as explained in Appendix \ref{spectrum}, one can, in this case, restrict our analysis to
$\lambda\geq 0$ and $h\geq 0$.

\begin{figure}[ht]
\includegraphics[width=100mm]{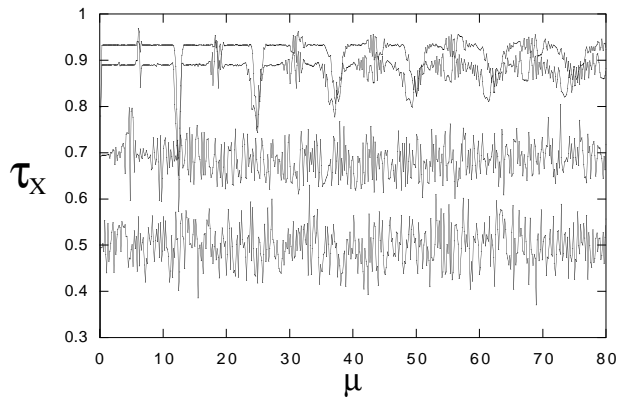}
\vspace{-15mm}
\caption{Behavior of $\tau_X$ as a function of $\mu=4 \lambda t/N$ for different values of the field $h= 0.5, 0.55, 0.8, 1$
from bottom to top ($\lambda=+1$ and $N=10^3$).}
\label{tauXt2}
\end{figure}

\begin{itemize}

\item For $h=0$, the $|X \rangle$ state is an eigenstate of $H$ so that $\tau_X(h=0)=0$ and $C_X(h=0)=0$ for all time $t$. 

\item For $\lambda=0$, $H$ is completely separable and thus both $\tau_X$ and $C_X$
vanish. However, for this state, this limit is very singular since, for any finite coupling $\lambda$, the large-field
limit leads to nontrivial entanglement properties ($\tau_X \rightarrow  1$) as can be seen in Fig. \ref{tauXt2}. 

\end{itemize}

%
%
%
%
%
%
\begin{figure}[ht]
\includegraphics[width=100mm]{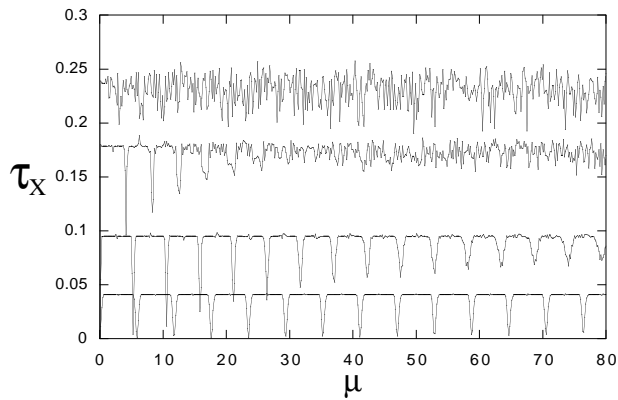}
\vspace{-15mm}
\caption{Same as Fig. \ref{tauXt1} for $h= 0.2, 0.3, 0.4,0.45$ from bottom to top.}
\label{tauXt1}
\end{figure}
%
%

Apart from these limiting cases, we have been unable to get analytical expressions of the one-tangle for this state.
We have displayed in Figs. \ref{tauXt2}  and  \ref{tauXt1} the behavior of $\tau_X$ for different values of the magnetic field 
as a function of
$\mu$. 

\begin{figure}[ht]
\includegraphics[width=100mm]{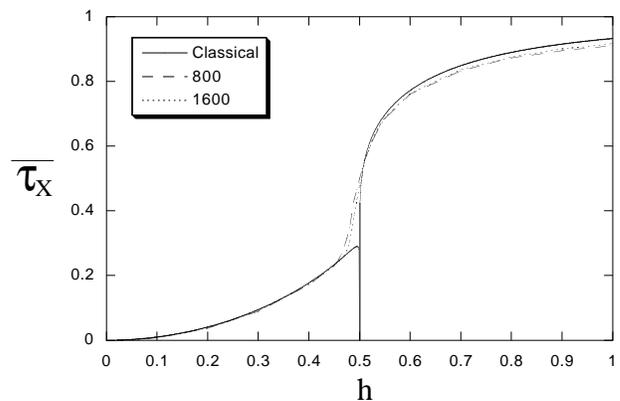}
\vspace{-15mm}
\caption{Time-averaged value of $\tau_X$ as a function of the magnetic field $h$ for $N=800,1600$ ($\lambda=+1$). The solid
line is obtained from the ``(semi) classical time-averaged one-tangle" as explained in the text.}
\label{tauXmean}
\end{figure}

For $h \ll \lambda$, the characteristic energy scale driving the dynamics is $\lambda/N$ and
hence $\tau$ is almost periodic in $\mu$ with period $2 \pi$ [Fig. \ref{tauXt1} (bottom)]. In this regime, quantum fluctuations
are weak and
$\tau$ is a smooth function of $\mu$. 
When $\mu$ increases, the depth of the hollows which corresponds to a revival of the
wave packet decreases and eventually goes to zero in the large $\mu$ limit so that, in this limit, $\tau_X$ is almost constant.
When $h$ increases, quantum fluctuations become more and more important and $\tau_X$ becomes more and more ``noisy."  The
maximum of these fluctuations is reached for $h=\lambda/2$ which corresponds to the critical field of the semi-classical
dynamics \cite{Milburn97,Micheli03}. Below (above) this critical field, $\langle S_x \rangle$ (which represents the difference
of population between the two wells in the BEC problem) oscillates around a nonvanishing (vanishing) value. 
In the other limit $h \gg \lambda$ where the magnetic field dominates $\tau_X$ goes to 1 and displays a
pseudo periodic behavior in $\mu$ with period $4 \pi$ as can already be inferred for $h=\lambda$ in Fig . \ref{tauXt2}
(top). At large times, the depth of the cusplike behavior present for
$\mu=0 [4\pi]$ decreases and $\tau$ is also almost constant. To go beyond this qualitative description, we have computed the
time-averaged value of the one-tangle $\overline{\tau_X}$ as a function of $h$. As shown in Fig. \ref{tauXmean},
the large $N$ limit of $\overline{\tau_X}$ can be partially captured by a simple semiclassical
approach, especially, when quantum fluctuations are weak; {\it i.~e.}, far from $h=\lambda/2$.
%
%
%
%

For a detailed discussion of this approach, we refer the reader to Refs. \cite{Milburn97,Micheli03}. The main idea is to
compute the solution of the semiclassical equations for the quantity $S_\alpha$. For an initial state polarized in the $x$
direction, these equations can be exactly solved following Eilbeck {\it et al.} \cite{Eilbeck85} and one gets
%
%
\begin{eqnarray}
S^{sc}_x(t) &=&{N\over 2} \:{\rm cn}\left( 2 h t|k^2 \right), \\
S^{sc}_y(t) &=&-{N\over 2}\: {\rm dn}\left( 2 h t|k^2 \right) \:{\rm sn}\left( 2 h t|k^2 \right), \\
S^{sc}_z(t) &=&{N \lambda \over 4h} \:{\rm sn}^2\left( 2 h t|k^2 \right),
\end{eqnarray}
%
%
where cn, dn, and sn are the Jacobi elliptic functions and where  $k=\lambda / 2h$  (see Appendix \ref{semiclassical} for
details). These $S^{sc}_\alpha$'s are periodic function with a period $2 K(k^2)/h$ for $k^2<1$ and $2 K(k^2)/ \lambda$ for
$k^2>1$ where $K$ denotes complete elliptic integral of first kind. Note that the case $k^2=1$ is singular since the upper
and lower limits do not coincide.  

Of course, if we compute the one-tangle using these semiclassical expressions, we trivially find
%
%
\begin{equation}
\tau_X^{sc}=1-{4 \over N^2} \sum_\alpha (S^{sc}_{\alpha})^2 =0
\end{equation}
%
%
and, consequently, $\overline{\tau_X^{sc}}=0$ for all $h$.
However, as said above, the fluctuations of $\tau_X$ are pretty weak at least away from $h=\lambda/2$. Thus, assuming that
$\tau_X$ is constant, we replace $\overline{\langle S_{\alpha} \rangle^2}$ by $\left(\overline{S_\alpha^{sc}}\right)^2$:
%
%
\begin{equation}
\overline{\tau_X}=1-{4 \over N^2} \sum_\alpha \overline{\langle S_{\alpha} \rangle^2} 
\simeq 1-{4 \over N^2} \sum_\alpha \left(\overline{S_\alpha^{sc}}\right)^2.
\label{approximation}
\end{equation}
%
%
Obviously, this crude approximation is to be justified only for constant $\langle S_{\alpha} \rangle$ and $S_\alpha^{sc}$,
provided the semiclassical description is meaningful; {\it i.~e.}, in the large-$N$ limit. 
After time-averaging the $S_\alpha^{sc}$'s over one period, we thus get a ``semiclassical time-averaged one-tangle." 
This quantity is depicted in Fig. \ref{tauXmean}, and as expected it is in excellent agreement away from $h=\lambda/2$.
Near this point, since the quantum fluctuations are very strong, the approximation (\ref{approximation}) fails and the
discrepancy is large. 
In the large-field limit, one has $\overline{S_\alpha^{sc}}=0$ for all $\alpha$ and thus  $\overline{\tau_X}=1$.\\

\begin{figure}[ht]
\includegraphics[width=100mm]{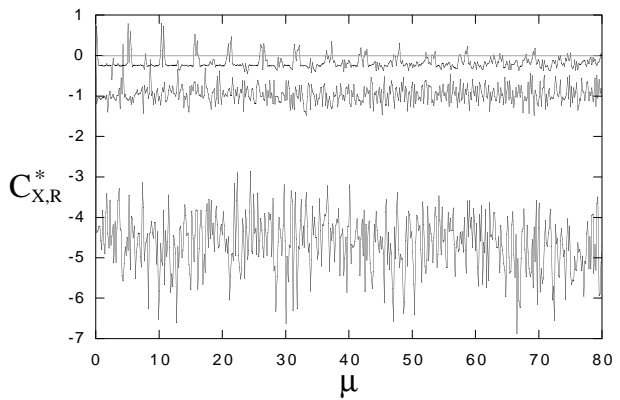}
\vspace{-15mm}
\caption{Behavior of $C^*_{X,R}$ as a function of $\mu=4 \lambda t/N$ for different values of the field 
$h= 0.3, 0.4, 0.45$ from top to bottom ($\lambda=+1$ and $N=10^3$). The grey line corresponds to the
threshold $C=0$.}
\label{concXt1}
\end{figure}

\begin{figure}[ht]
\includegraphics[width=100mm]{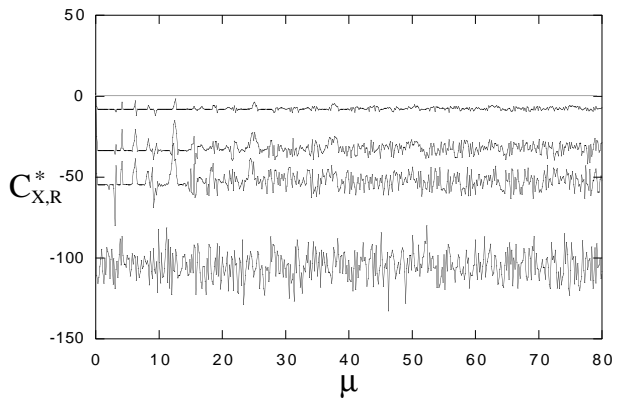}
\vspace{-15mm}
\caption{Same as Fig. \ref{concXt1} for $h= 0.5, 0.8, 1,2$ from bottom to top.}
\label{concXt2}
\end{figure}

Let us now discuss the concurrence dynamics for the $|X\rangle$  state whose behavior is depicted in Figs. \ref{concXt1} and 
\ref{concXt2} for different values of the magnetic field. At small field, [$h < \lambda/2$ (see Fig. \ref{concXt1})],
$C^*_{X,R}$ is most often negative but displays some peaks (where $C^*_{X,R}>0$) which coincides with the hollows in $\tau_X$
and whose amplitude goes to zero when $\mu$ increases. As a result, at large times, the rescaled concurrence $C_{X,R}$ always
vanishes. When $h$ increases, this phenomenon is amplified and, as for the one-tangle, $C^*_{X,R}$ fluctuates more and more until
$h=\lambda/2$ where it reaches its minimum value. Above this critical field, $C^*_{X,R}$ increases when $h$ increases, but
remains always negative at large times.  Note that at short times there is, for any $h$, a ``peak of concurrence" for which
$C^*_{X,R}$ is positive but as time goes on, it is always negative. 

To summarize, at sufficiently large times and for any values of the magnetic field $h$, the rescaled concurrence
$C_{X,R}$ always vanishes.

%
%
%
%
%
%
%
%
 
%
%
%
\subsection{The $|Z \rangle$ state}
\label{Zstate}
%
%
The $|Z \rangle$ state is an eigenstate of the operator $\prod_i \sigma_z^i$ which commutes with $H$. As a consequence, at
all times, one has $\langle S_x \rangle=\langle S_y \rangle=0$, so that 

%
%
\begin{equation}
\tau_Z=1-{4 \over N^2} \langle S_{z}\rangle^2.
\end{equation}
%
%

In addition and contrary to the $|X \rangle$ state, it is
obvious that the direction of the magnetic field $h$ plays a role in the dynamics of $|Z \rangle$. Thus, to investigate all
possible situations, we consider the case $\lambda\geq 0$ for all $h$, the region $\lambda < 0$ being obtained
by transformations discussed in Appendix \ref{spectrum}.

\begin {itemize}

\item For $h=0$, $H$ is trivially solvable as well as the time evolution of the $|Z \rangle$ state. As detailed in Ref.
\cite{Wang02}, one finds, in this case

%
%
\begin{eqnarray}
\tau_Z(h=0,t) &=&1-\cos(\mu/2)^{2(N-1)}, \\
C_{Z,R}(h=0,t) &=& {1\over 4} \Bigg\{\cos(\mu)^{N-2}-1+ \nonumber\\
&& \Big(\big[\cos(\mu)^{N-2}-1 \big]^2  +\\
\label{concZ0}
&& \big[ 4 \cos(\mu/2)^{N-2} \sin(\mu/2) \big]^2 \Big)^{1/2} \Bigg\} \nonumber,
\end{eqnarray}
%
%
where, as previously,  $\mu=4 \lambda t/N$. The expression for $C_{Z,R}(h=0,t)$ given above is
obtained by noting that the condition 
%
%
\begin{equation}
2\rho^{(2)}_{22}< \sqrt{\rho^{(2)}_{11} \rho^{(2)}_{44}} + \big|\rho^{(2)}_{14}\big|
\end{equation}
%
%
is always satisfied. Thus in the large-$N$ limit, $\tau_Z$ is equal to 1 except for $\mu=0 \: [2 \pi]$ where
$\tau_Z=0$ and, as expected, the concurrence vanishes for all $\mu$. However, if one considers the rescaled concurrence, a
close inspection of expression (\ref{concZ0}) shows that a periodic pulse also occurs for $\mu=0 \: [2 \pi]$ whose width
becomes sharper and shaper as $N$ increases and whose maximum height goes to 1 in the large-$N$ limit. 

At this stage, it is interesting to note that $\tau_Z$ and $C_{Z,R}$ are strongly correlated in the sense that, when the
one-spin entanglement is large (and here maximum), the two-spin entanglement is small except for 
$\mu\simeq 0 \: [2 \pi]$. This is a relatively surprising result which is also encountered in other regimes (see below).
Indeed, usually, the trace operation destroys some correlations and one expects that the absence of two-spin entanglement
might not be able to generate one-spin entanglement. Nevertheless, as discussed in Sec. \ref{Measures}, this phenomenon is
met for the $|{\rm GHZ} \rangle$ state, but here it is encountered for a nonstationary state which preserves this
feature as time goes on. \\

\item For $\lambda=0$, the $|Z \rangle$ state is an eigenstate of $H$ so that $\tau_Z(\lambda=0)=0$ and $C_Z(\lambda=0)=0$ for
all time $t$. Contrarily to the $|X \rangle$ state, the limit $\lambda=0$ coincides with the high field regime at finite
$\lambda$. Indeed, as discussed in Ref. \cite{Vidal04_1},  the $|Z \rangle$ state is the
ground state of $H$ for $h>\lambda$ in the thermodynamical limit, so that $\tau_Z(h>\lambda) \simeq 0$ and
$C_Z(h>\lambda)\simeq 0$.

However, the rescaled concurrence has a nontrivial behavior that can be captured due to its relation with the spin squeezing
parameter \cite{Kitagawa93}
%
%
\begin{equation}
\xi^2= {4(\Delta S_{\vec{n}_\perp})^2 \over N},
\label{spinsqueezingdef}
\end{equation}
%
%
which measures the spin fluctuations of a correlated quantum state \cite{Kitagawa93}.  
The subscript $\vec{n}_\perp$ refers to an axis perpendicular to the mean spin $\langle \vec{S} \rangle$ where
the minimal value of the variance is obtained.

Indeed, as shown in Refs. \cite{Messikh03,Wang03}, for any state belonging to the subspace $S=N/2$ and eigenstate of
the operator $\prod_i \sigma_z^i$, one has 
%
%
\begin{equation}
\xi^2=1-C_R,
\label{squeezing}
\end{equation}
%
%
provided the reduced density matrix elements of the state considered verify $|\rho^{(2)}_{14}|\geq \rho^{(2)}_{22}$. 
We have checked that actually this inequality is always satisfied for any $h$ and for all the times under investigation. 

\item The relation (\ref{squeezing}) thus allows us to deduce the rescaled concurrence using the frozen-spin approximation 
used in Ref. \cite{Law01} to compute $\xi^2$ in two different regimes.

\begin{itemize}

\item For $h>\lambda$, as explained above, the $|Z \rangle$ state is the ground state of $H$ in the infinite $N$ limit. So,
even at finite $N$, one expects $\langle S_z \rangle$ to be almost constant in this region. The so-called frozen-spin
approximation precisely consists in replacing, in the Heisenberg equation, $S_z(t)$ by its initial value $N/2$. Then,
following \cite{Law01} and using Eq. (\ref{squeezing}), one gets an analytical expression for the
rescaled concurrence,
%
%
\begin{equation}
C_{Z,R}=1-\bigg[\cos^2(\omega t) +{w^2 \over 4 h^2 } \sin^2(\omega t) \bigg],
\label{frozen1}
\end{equation}
%
%
with $\omega=2\sqrt{h(h-\lambda)}$.\\

\item For $h<0$, the situation is, in fact, similar. Indeed, as shown in Ref.
\cite{Vidal04_2}, the
$|Z
\rangle$ state is the ground state of $H(\lambda<0, h>0)$ in the thermodynamical limit. As detailed in Appendix \ref{spectrum},
this implies that it is the highest-energy state of $H(\lambda>0, h<0)$. Thus, the frozen-spin approximation can
also be used in the region $( h< 0)$ and, according to \cite{Law01}, one then obtains
%
%
\begin{equation}
C_{Z,R}=1-\bigg[\cos^2(\omega t) + {4 h^2 \over w^2} \sin^2(\omega t) \bigg].
\end{equation}
%
%
\end{itemize}

Of course, this approximation is more and more valid in these regions ($h<0$ and $h>\lambda$) when $N$ increases.
Conversely, at fixed $N$, it is safe for either $h \gg \lambda$ or $h \ll 0$.

\end{itemize}

\begin{figure}[ht]
\includegraphics[width=100mm]{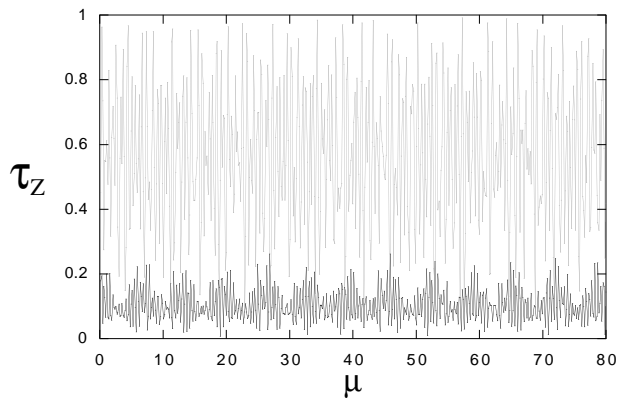}
\vspace{-15mm}
\caption{Behavior of $\tau_Z$ as a function of $\mu=4 \lambda t/N$ for $h=0.01$ (grey) and $h=0.9$ (black) 
($\lambda=+1$ and $N=10^3$).}
\label{tauZt}
\end{figure}

The only region where the $|Z \rangle$ state is not an eigenstate of $H$ in the thermodynamical limit is $0< h < \lambda$. 
We have plotted in Fig. \ref{tauZt} the behavior of $\tau_Z$ as a function of the rescaled dimensionless time $\mu=4 \lambda
t/N$ for different values of the magnetic field $h$ in this range. 
%
%
%
%

\begin{figure}[ht]
\includegraphics[width=100mm]{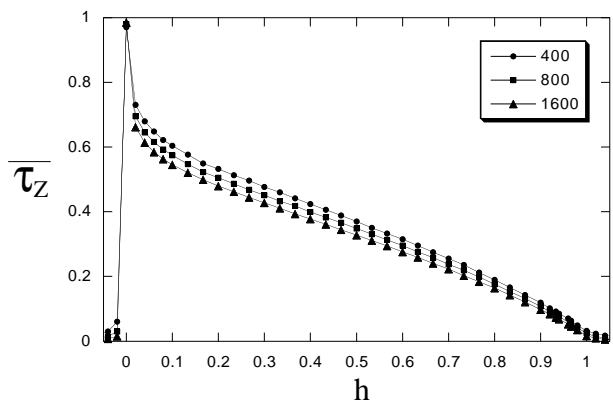}
\vspace{-15mm}
\caption{Time-averaged value of $\tau_Z$ as a function of the magnetic field $h$ for $N=400,800,1600$ ($\lambda=+1$).}
\label{tauZmean}
\end{figure}

As explained above, $\tau_Z \simeq 0$ for  $h<0$ and $h>\lambda$ since, there, $|Z \rangle$ is an eigenstate of $H$ in the
large-$N$ limit. By contrast, for $h\in [0,\lambda]$, $\tau_Z$ is an oscillating function of $\mu$ with an average value that
is nonvanishing as can be seen in Fig. \ref{tauZt}. 
The time-averaged value of $\tau_Z$ for different $N$ is shown in Fig. \ref{tauZmean}. 
%
%
%
%

\begin{figure}[ht]
\includegraphics[width=100mm]{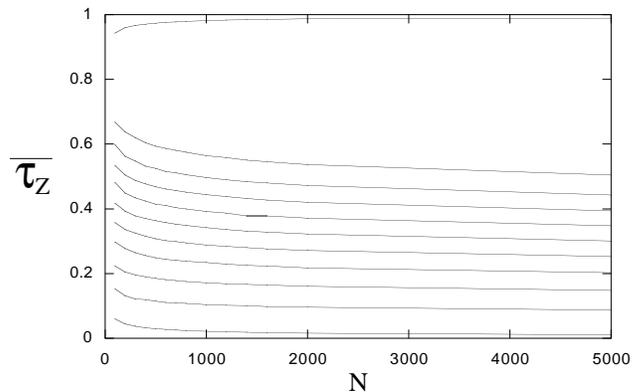}
\vspace{-15mm}
\caption{Behavior of $\overline {\tau_Z}$ as a function of $N$  for $h=0$ (top) to 1 (bottom) with a step
$\Delta h=10^{-1}$ ($\lambda=+1$).}
\label{tauZmeanN}
\end{figure}

As can be seen, $\overline{\tau_Z}$
is a decreasing function of $N$ at fixed $h$ except for $h=0$ as discussed above. To determine its large-$N$ behavior, 
we have computed $\overline{\tau_Z}$ for several $h$, as a function of $N$, up to $N=5000$.  The results displayed in Fig.
\ref{tauZmeanN} do not allow us to extract an asymptotic value but it is likely that for $h\in [0,\lambda[$, 
$\lim_{N\rightarrow \infty} \overline{\tau_Z} \neq 0$. 
%
%
%
%

As for the $|X \rangle$ state, we could have tried to reproduce this behavior by a classical analysis, but unfortunately, the
$|Z \rangle$ state is a fixed point of the classical equations of motion (\ref{eqa1}) and (\ref{eqtheta}). Thus, one must in this
case use a real semiclassical treatment analogous to that presented in \cite{Micheli03} which consists in averaging over classical
trajectories near this fixed point. However, as shown in Fig. \ref{tauZt}, $\tau_Z$ is a strongly oscillating function so that
the approximation (\ref{approximation}) is not valid and we did not compute the ``semiclassical time-averaged
one-tangle" for the $|Z \rangle$ state.

\begin{figure}[ht]
\includegraphics[width=100mm]{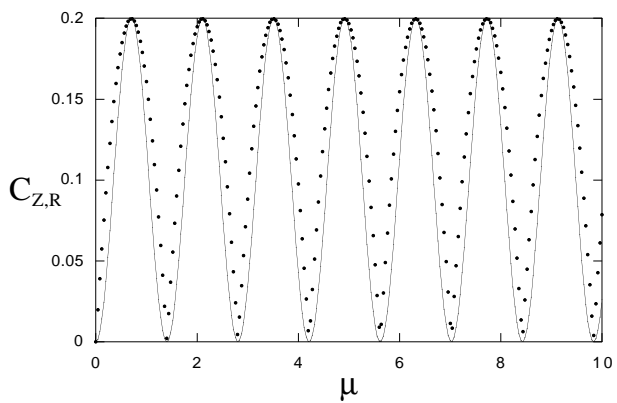}
\vspace{-15mm}
\caption{Behavior of $C_{Z,R}$ as a function of $\mu=4 \lambda t/N$ for $h=5$ ($\lambda=+1$ and $N=10^3$).
Dots are obtained from the exact numerical diagonalizations and the solid line is computed with the frozen-spin
approximation (\ref{frozen1}).}
\label{concZt5}
\end{figure}

\begin{figure}[ht]
\includegraphics[width=100mm]{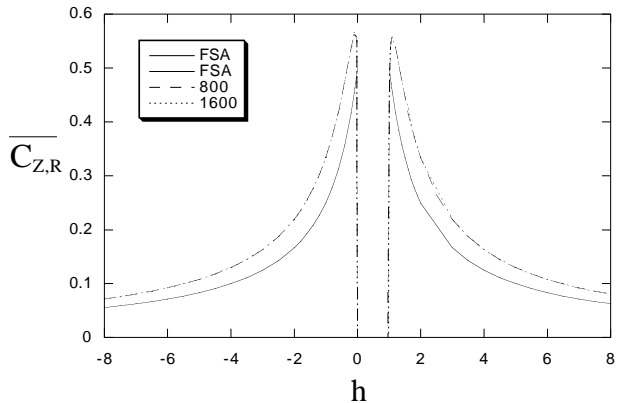}
\vspace{-15mm}
\caption{Time-averaged value of $C_{Z,R}$ as a function of the magnetic field $h$  for $N=800,1600$ ($\lambda=+1$). The solid
line corresponds to both branches ($h<0$ and $h>1$) of the frozen-spin approximation (FSA).}
\label{concZmean}
\end{figure}

We now focus on the concurrence dynamics for the $|Z \rangle$ state, whose behavior is shown in Fig. \ref{concZt5} for
$h=5 \lambda$. As can be seen, the frozen-spin approximation is, in this case, in good agreement with the exact
(numerical) results.
%
%
%
%
For $h\in [0,\lambda]$, $C^*_{Z,R}$ is always negative so that $C_{Z,R}=0$. It is worth noting that there is once again an
``anticorrelation" between the one-tangle and the rescaled concurrence since $\tau_X\neq 0$ when $C_{Z,R}=0$ for $h\in
[0,\lambda]$, whereas for $h\in ]-\infty,0] \cup [\lambda, \infty[$, one has $\tau_X=0$ and $C_{Z,R}\neq 0$. Indeed, in this
latter region where $|Z \rangle$ becomes an eigenstate in the thermodynamical limit, one always has a non vanishing
rescaled concurrence as can be seen in Fig. \ref{concZmean} where we have displayed the time-averaged value
of $C_{Z,R}$ as a function of $h$. At large $|h|$ the frozen-spin approximation allows us to extract the asymptotic behavior 
$\overline{C_{Z,R}}\sim |\lambda/h|$.
%
%
%
%

%
%
%
\section{Conclusion and outlooks}
%
%
%
We have analyzed the one-spin and two-spin entanglement dynamics for two different states. In the case of a state fully
polarized in the $x$ direction, we have found a nontrivial behavior of the one-tangle whose time-averaged value increases
monotonically when the magnetic field increases whereas it has a vanishing concurrence for all $h$. By contrast, for an initial
state fully polarized in the $z$ direction, the one-tangle is nonvanishing only for $h\in [0,\lambda]$  where its
concurrence vanishes. Apart from this region, the one-tangle vanishes in the thermodynamical limit (since $|Z \rangle$ is then
an eigenstate of $H$) but the rescaled concurrence has a nontrivial time-averaged value which can be understood using the
frozen-spin approximation introduced in Ref. \cite{Law01}. 

Contrary to the ground-state entanglement, we did not find a nontrivial singularity at the critical point $h=\lambda$ but only
some enhanced fluctuations near $h=\lambda/2$ for the $| X \rangle$ state. For the  $| Z \rangle$ state, we have not detected
any peculiarity at this point except that $\partial_h \overline{\tau_Z}$ seems to be minimum there whereas $\partial_h
\overline{\tau_X}$ is rather maximum.

We wish to emphasize that some experiments may be possible to investigate these dynamical entanglement, especially the
concurrence. Indeed, as explained in the Sec. \ref{Zstate}, the rescaled concurrence can be related to the spin squeezing
parameter as established in Refs. \cite{Wang03,Messikh03}. Thus, the behavior of the $C_R$ as a function of the field
should, in principle, be measured for both the ground state and $|Z(t)\rangle$. Nevertheless, such an exciting perspective is
of course  conditioned by the possibility to measure the squeezing parameter. 
Bose-Einstein condensates are certainly the best candidates for such experiments since it has already been possible to measure
the angular momentum in such systems \cite{Chevy00}.


\acknowledgments

We would like thank B. Dou\c{c}ot, J. Dukelsky, R. Mosseri, and D. Mouhanna for fruitful discussions.
%
%
%
%
\appendix
%
%
\section{Spectrum symmetries and their implications for the dynamics}
\label{spectrum}
%
%
\label{symmetries}
Let us consider an eigenstate of the Hamiltonian $H$,
%
%
\begin{equation}
|\phi\rangle=\sum_{i=0}^{2^N-1} \alpha_i |i \rangle,
\end{equation}
%
%
such that 
%
%
\begin{equation}
H(\lambda,h) |\phi\rangle= E |\phi \rangle
\end{equation}
%
%
it is then straightforward to show that
%
%
\begin{equation}
|\varphi \rangle=\sum_{i=0}^{2^N-1} \alpha_{2^N-1-i} |i \rangle,
\end{equation}
%
%
satisfies
%
%
\begin{eqnarray}
H(-\lambda,+h) |\varphi \rangle&=& -E |\varphi \rangle,\\
H(\lambda,-h) |\varphi \rangle&=& E |\varphi \rangle.
\end{eqnarray}
%
%
Here, we have used a natural coding of the Hilbert space states where the number $i$ (and its associated state $|i \rangle$)
is in a one-to-one correspondence with its binary decomposition expressed in terms of $\uparrow$ (1) or
$\downarrow$ (0) spins. 
These identities show that the spectrum of $H$ is odd under the transformation $\lambda \rightarrow -\lambda$
and even under $h \rightarrow -h$. Moreover, they also imply that
%
%
\begin{equation}
H(-\lambda,-h) |\phi \rangle= -E |\phi \rangle,
\label{symspec}
\end{equation}
%
%
so that $H(+\lambda,+h)$ and $H(-\lambda,-h)$ have the same eigenstates but with opposite eigenenergies.

Let us now consider a state $|\psi\rangle$ whose spectral decomposition on the eigenstates of $H$ reads
%
%
\begin{equation}
|\psi\rangle=\sum_{j=0}^{2^N-1} a_j |\phi_j \rangle,
\end{equation}
%
%
where all the $a_i$'s are real numbers, and an observable $A$ satisfying
%
%
\begin{equation}
\langle \phi_{j'}|A |\phi_j \rangle =\langle \phi_j|A |\phi_{j'} \rangle,
\end{equation}
%
%
for all $(j,j')$. Then from Eq. (\ref{symspec}) one has
%
%
\begin{equation}
\langle \psi_{+,+}(t)|A| \psi_{+,+}(t) \rangle =\langle \psi_{-,-}(t)|A| \psi_{-,-}(t) \rangle,
\end{equation}
%
%
where $| \psi_{\pm ,\pm}(t) \rangle=e^{-i H(\pm \lambda,\pm h) t} |\psi\rangle$. In addition, it is clear that the density
matrix matrices $\rho_{+,+}$ and $\rho_{-,-}$ are identical.

%
%
%
\section{Semiclassical approach}
\label{semiclassical}
%
%
%
The Heisenberg equations for the spin operators are given by
%
%
\begin{eqnarray}
\dot S_x&=& 2 h S_y,\\
\dot S_y&=&-2 h S_x+{2 \lambda \over N} ( S_z S_x+ S_x S_z ),\\
\dot S_z&=&-{2 \lambda \over N} ( S_y S_x+ S_x S_y ).
\end{eqnarray}
%
%
These equations cannot be solved exactly for any $(\lambda,h)$ and any initial conditions. To investigate the spin dynamics,
we follow Milburn {\it et al.} \cite{Milburn97} and consider the semiclassical dynamics (mean-field approximation) associated with
these equations by considering the ``double-well" parametrization (Schwinger-like) for the spin variables which reads
%
%
\begin{eqnarray}
s_x&=& {N\over 2} \left( |b_2|^2-|b_1|^2       \right) , \\
s_y&=&-{iN\over 2} \left( b_1^* b_2 - b_1 b_2^* \right) , \\
s_z&=&{N\over 2}  \left( b_1^* b_2 + b_1 b_2^* \right) .
\end{eqnarray}
%
%
The complex variables $b_1$ and $b_2$ satisfy the constraint $|b_1|^2+|b_2|^2=1$ and obey
%
%
\begin{equation}
\dot b_j=i h b_{3-j} +2 i \lambda |b_j|^2 b_j,
\label{nonlinear}
\end{equation}
%
%
for $j=1,2$.
This type of equation has been studied in detail by Eilbeck {\it et al.}  so that we only sketch here the main
lines of the solutions and refer the reader to Ref. \cite{Eilbeck85} for details.
Setting $b_j=a_j e^{i \theta_j}$ (for $j=1,2$) and $\theta=\theta_1-\theta_2$, the latter nonlinear equation system
(\ref{nonlinear}) is rewritten as
%
%
\begin{eqnarray}
\dot a_1 &=& h \left( 1-a_1^2 \right)^{1/2} \sin \theta  \label{eqa1} ,\\
\dot \theta &=& 2 \left( a_1^2 -1 \right) \left( 2\lambda -{h \cos  \theta \over a_1 \left( 1-a_1^2 \right)^{1/2}}\right).
\label{eqtheta}
\end{eqnarray}
%
%
Then, using the conservation of energy,
%
%
\begin{eqnarray}
E(a_1,\theta)&=& - \lambda \left[ 1- 2a_1^2 \left(1- a_1^2 \right) \right]  \label{conservation} \\
&&-2h a_1 \left( 1-a_1^2\right)^{1/2} \cos \theta,
\nonumber \\ 
&=&{1\over N} \left[ H(a_1,\theta)-\lambda/2 - {\lambda N \over 2} \right],
\end{eqnarray}
%
%
we can eliminate $\theta$ to get a closed equation for $a_1$.

From now on, we consider the initial conditions $a_1(0)=1$ and $\theta(0)=0$ which describe a state polarized in
the $-x$ direction; i.e., the classical counterpart of the $|-X \rangle$ state. This state is not the one considered in the
Sec. \ref{xstate} since it lies in the opposite direction, but for clarity, we have used the same notation as in Refs.
\cite{Milburn97,Eilbeck85}. One thus has 
%
%
\begin{equation}
E(a_1,\theta)= - \lambda,
\end{equation}
%
%
so that from the energy conservation (\ref{conservation}), one gets
%
%
\begin{equation}
\cos \theta={\lambda \over h} a_1(1-a_1^2)^{1/2}.
\end{equation}
%
%
Next, using Eq. (\ref{eqa1}) and setting $a_1^2=(1+y)/2$, it is straightforward to show that
%
%
\begin{equation}
\left({\dot y \over 2h}\right)^2=1-k^2+(2k^2-1) y^2-k^2 y^4,
\end{equation}
%
%
where $k=\lambda/(2h)$. The solution of this equation is the Jacobi elliptic function
%
%
\begin{equation}
y(t)={\rm cn} (2h t|k^2),
\end{equation}
%
%
which yields
%
%
\begin{eqnarray}
S_x(t) &=&{N\over 2} \left( 1-2a_1^2 \right)=-{N\over 2} \:{\rm cn}\left( 2 h t|k^2 \right) ,\\
S_y(t) &=&-N a_1 \dot a_1/h={N\over 2} \:{\rm dn}\left( 2 h t|k^2 \right) \:{\rm sn}\left( 2 h t|k^2 \right) ,\hspace{10mm}\\
S_z(t) &=&{N \lambda \over h} a_1^2 (1-a_1)^2 ={N \lambda \over 4h} \:{\rm sn}^2\left( 2 h t|k^2 \right).
\end{eqnarray}
%
%

Finally, since we are not interested in the $|-X \rangle$ but in the $|X \rangle$ state, we just have to change the sign
correctly and we get the expression given in Sec. \ref{xstate}.

%
%
\section{One-spin  and two-spin reduced density matrices of fully symmetric states}
%
%
Let us consider a state 
%
%
\begin{equation}
|\psi\rangle=\sum_{M=-N/2}^{+N/2} \alpha_M |N/2,M\rangle,
\end{equation}
%
%
where $\{|S,M\rangle \}$ is an eigenbasis of ${\bf S}^2$ and $S_z$, and its density matrix $\rho=|\psi\rangle \langle\psi |$.
Here, we restrict our discussion to states belonging to the maximum spin sector $S=N/2$ which are relevant for BEC. 

In this subspace, all states are invariant under the permutation group ${\cal S}_N$, so that one can easily compute the  matrix
elements of the reduced density matrices $\rho_i^{(1)}$ and $\rho_{i,j}^{(2)}$.  
Note that the permutation symmetry implies that $\rho_i^{(1)}$ does not depend on $i$, and  that $\rho_{i,j}^{(2)}$ does not
depend on $i$ and $j$, so that we omit these indices in the following.\\
 
As is well known, the one-spin reduced density matrix is easily expressed in terms of the one-spin correlation functions
or, more precisely, in terms of the $\langle \sigma_{\alpha} \rangle$'s. In the eigenbasis
$\{|\uparrow \rangle, |\downarrow \rangle\}$ of $\sigma_{z}$, a straightforward calculations gives
%
%
\begin{eqnarray}
\rho^{(1)}_{11}&=& {1 \over 2} + \sum_M |\alpha_M|^2 {M \over N}, \\
\rho^{(1)}_{22}&=& {1 \over 2} - \sum_M |\alpha_M|^2 {M \over N}, \\
\rho^{(1)}_{12}&=& \sum_M \alpha_M \alpha_{M+1}^* {\sqrt{(N-2M+2) (N+2M)}\over 2N}, \:\:\:\:\:
\end{eqnarray}
%
%
with $\rho^{(1)}_{ij}=\rho^{(1)*}_{ji}$.

The two-spin reduced density matrix can also be easily expressed in terms of the two-spin correlation functions
(see Ref. \cite{Wang03}). In the eigenbasis 
$\{|\uparrow \uparrow \rangle, |\uparrow \downarrow \rangle,|\downarrow \uparrow \rangle,|\downarrow \downarrow \rangle \}$,
one has
%
%
\begin{eqnarray}
\rho^{(2)}_{11}&=& \sum_M |\alpha_M|^2 {(N+2M)(N+2M-2) \over 4N(N-1)}, \\
\nonumber\\
\rho^{(2)}_{12}&=& \sum_M \alpha_M \alpha_{M+1}^* (N+2M-2)  \nonumber\\
& &\times {\sqrt{(N-2M+2) (N+2M)} \over 4N(N-1)}, \\
\nonumber\\
\rho^{(2)}_{14}&=& \sum_M \alpha_M \alpha_{M+2}^* \sqrt{(N+2M)(N+2M-2)}  \nonumber\\
& &\times {\sqrt{(N-2M+2)(N-2M+4)} \over 4N(N-1)},\\
\nonumber\\
\rho^{(2)}_{24}&=& \sum_M \alpha_M \alpha_{M+1}^* (N-2M)  \nonumber\\
& & \times {\sqrt{(N-2M+2) (N+2M)} \over 4N(N-1)}, \\
\nonumber\\
\rho^{(2)}_{22}&=& \sum_M |\alpha_M|^2 {(N-2M)(N+2M) \over 4N(N-1)}, \\
\nonumber\\
\rho^{(2)}_{44}&=& \sum_M |\alpha_M|^2 {(N-2M)(N-2M-2) \over 4N(N-1)},
\end{eqnarray} 
%
%
with $\rho^{(2)}_{13}=\rho^{(2)}_{12}$, $\rho^{(2)}_{23}=\rho^{(2)}_{33}=\rho^{(2)}_{22}$, 
$\rho^{(2)}_{34}=\rho^{(2)}_{24}$, and $\rho^{(2)}_{ij}=\rho^{(2)*}_{ji}$.


\end{document}